\documentclass[aps,12pt,preprint,showpacs]{revtex4}

\usepackage{graphicx}
\usepackage{bm}
\usepackage{amsmath}

\newcommand{\NITEROI}{Instituto de F\'{\i}sica da Universidade 
Federal Fluminense, Boa Viagem 24210-340, 
Niter\'oi - RJ, Brazil.}
\newcommand{\INMETRO}{Instituto Nacional de Metrologia,
  Normaliza\c{c}\~ao e Qualidade Industrial, Rua Sta. Alexandrina 416,
  Rio Comprido 20261-232, RJ, Brazil.}
\newcommand{\UQ}{ARC Centre of Excellence for Quantum-Atom Optics,
  School of Physical Sciences, University of Queensland, Brisbane, Qld
4072, Australia.}

\newcommand{\epl}{Europhys. Lett.\ }

\begin{document} 
\title{Quantum phase-space analysis of the pendular cavity.} 
\author{M.~K. Olsen$^{1,2}$, A.~B. Melo$^{1,3}$, K. Dechoum$^{1}$ and A.~Z. Khoury$^{1}$.}    
\affiliation{$^{1}$\NITEROI\\
             $^{2}$\UQ\\
             $^{3}$\INMETRO}

\date{\today}
 
\begin{abstract} 
We perform a quantum mechanical analysis of the pendular cavity, using the positive-P representation, showing that the quantum state of the moving mirror, a macroscopic object, has noticeable effects on the dynamics. This system has previously been proposed as a candidate for the quantum-limited measurement of small displacements of the mirror due to radiation pressure, for the production of states with entanglement between the mirror and the field, and even for superposition states of the mirror. However, when we treat the oscillating mirror quantum mechanically, we find that it always oscillates, has no stationary steady-state, and exhibits uncertainties in position and momentum which are typically larger than the mean values. This means that previous linearised fluctuation analyses which have been used to predict these highly quantum states are of limited use. 
We find that the achievable accuracy in measurement is far worse than the standard quantum limit due to thermal noise, which, for typical experimental parameters, is overwhelming even at $2$\;mK.       
\end{abstract}
\pacs{42.50.Lc,42.50.-p,42.50.Pq,03.65.Yz}
\maketitle

\section{Introduction}

The pendular cavity, or Fabry-Perot cavity in which one of the mirrors is free to oscillate, has previously been investigated by a number of researchers, both experimentally~\cite{Dorsel,Schiller,Numata}, and theoretically~\cite{Claude,nariz,Kurt,escroto,Bose,Imperial,Pinard,Vittorio,Vitali,Courty}. Closely related schemes have been theoretically proposed to entangle mirrors~\cite{camerino} and create superposition states of a mirror~\cite{nakedemperor}. Common to almost all the theoretical treatments is a linearisation of quantum Langevin equations around their classical steady-state solutions~\cite{Danbook}. This then allows for the relatively simple calculation of spectral quantities which may be measured outside the cavity. Results obtained in this way have been used to analyse the sensitivity of gravity wave interferometers~\cite{Pace}, predict the suppression of quantum noise~\cite{Claude,nariz}, propose quantum nondemolition measurements of photon number~\cite{Kurt}, analyse the quantum limits to measurements with an atomic force microscope~\cite{escroto}, analyse the quantum noise in position measurements of the oscillating mirror~\cite{Imperial}, calculate the phase noise in the cavity field~\cite{Vittorio}, predict the entanglement of macroscopic oscillators via radiation pressure~\cite{camerino}, and propose the quantum locking of interferometer mirrors~\cite{Courty}. Using the state-vector approach so common in quantum computing theory, it has been proposed that quantum superpositions, entanglement and near-number states of the cavity field, along with superposition states of the mirror, can be produced with this system~\cite{Bose}. Using a similar state-vector approach, it has been proposed that quantum superpositions of a mirror may be created by the interaction with a single photon~\cite{nakedemperor}.

It is well-known that the linearised fluctuation analysis used in the majority of the theoretical papers cited above is limited in its applicability. It has been shown, for example, that the mean-field equations derived in this way can give misleading results for travelling-wave second harmonic generation~\cite{OCrevive,revive,QND} and for the intracavity interaction between light and condensed atoms~\cite{PPcav,PPcavmais,Bratislava} as well as in Raman photoassociation of atomic Bose-Einstein condensates~\cite{JJHletter,JJH,RCstat,PRAstat}. The spectra calculated via this method are also known to not be accurate near any critical points of the system, as has been shown with the optical parametric oscillator~\cite{OPO1,OPO2}. There are two conditions which must be fulfilled for a linearised analysis to be trustworthy. The first has to do with the sign of the real part of the eigenvalues of the drift matrix of the equations written for the fluctuations; if these have the wrong sign the fluctuations can grow exponentially and the analysis loses its validity. The second has to do with the size of the fluctuations themselves, in relation to the classical steady-state values. As we will show, this second condition is not fulfilled for this system, as the quantum state of the mirror, or more accurately, of the mirror phonons, is reasonably expected to be thermal. A characteristic of a thermal state is that the variance is larger than the mean value, which therefore makes any expression of the mirror phonons as having some well-defined classical mean value plus small fluctuations rather dubious. Even though cooling of the mirror via feedback mechanisms has been achieved~\cite{exfeed}, and analysed theoretically~\cite{Ribichini}, due to the nature of the coupling between the electromagnetic field and the mirror phonons, it seems that all that can be achieved is a thermal state at a lower temperature, so that the problem remains.    

To treat the macroscopic mirror quantum mechanically we will begin with the Hamiltonian approach of Law~\cite{Law}, in the approximation that only a single optical mode is important, and extended to include cavity pumping and damping. To treat the fluctuations of the mirror which result from its coupling to a thermal reservoir, we will use the Brownian motion master equation developed by Di\'osi~\cite{Diosi}, which is suitable for the temperatures we will consider here. 
Following a common procedure in quantum optics~\cite{Crispin}, we will
develop a Fokker-Planck equation in the positive-P representation~\cite{P+}. This Fokker-Planck equation allows us to write stochastic differential equations which are an exact mapping from the system master equation, and which can be used to calculate any desired normally-ordered operator moments.

\section{System and Hamiltonian}
\label{sec:Ham}

We consider a system of a pumped Fabry-Perot cavity in which one of the mirrors is free to undergo oscillatory motion due to both the light pressure and thermal fluctuations. We use the standard annihilation, $\hat{a}$, and creation, $\hat{a}^{\dag}$ operators for the electromagnetic field, and the operators $\hat{x}$ and $\hat{p}$ for the displacement from the equilibrium position and the momentum of the mirror, which will be treated as a harmonic oscillator.
Neglecting the coupling of the mirror to its bath for the moment, we can write the Hamiltonian as 
\begin{equation}
\hat{H}=\hat{H}_{free}+\hat{H}_{int}+\hat{H}_{pump}+\hat{H}_{bath},
\end{equation}
where
\begin{eqnarray}
\hat{H}_{free} &=& \hbar\omega_{0}\hat{a}^{\dag}\hat{a}+\frac{1}{2m}\hat{p}^{2}
+\frac{1}{2}m\omega_{m}^{2}\hat{x}^{2},\nonumber\\
\hat{H}_{int} &=& -\hbar g\hat{a}^{\dag}\hat{a}\hat{x},\nonumber\\
\hat{H}_{pump} &=& i\hbar\epsilon(\hat{a}^{\dag}\mbox{e}^{i\omega_{0}t}-\hat{a}\mbox{e}^{-i\omega_{0}t})\nonumber\\
\hat{H}_{bath}&=&\Gamma\hat{a}^{\dag}+\Gamma^{\dag}\hat{a}.
\label{eq:hamiltoniano}
\end{eqnarray}
In the above, $\omega_{0}$ is the field frequency, $\epsilon$ represents the classical real pump,
$m$ is the mass of the mirror,
$\omega_{m}$ is the mirror oscillation frequency and $g=\omega_{0}/L$ is the coupling between the
mirror and the cavity field, with $L$ being the length of the cavity. The $\Gamma$s represent optical bath
operators. The damping of the mirror, which we will treat as Markovian, will be included at the next step.

We now wish to write a master equation for the density matrix of our combined system in a frame
rotating at $\omega_{0}$. To do this, we will make two different, but
consistent approximations for the damping of the cavity and the
mirror. The cavity reservoir will be considered to be at zero
temperature, which is consistent with the very high temperatures
necessary to produce thermal photons at the frequencies involved. The
mirror reservoir will be treated as being at a finite temperature,
which is necessary because of the number of thermal phonons which will
be present in the system. As the temperatures required to create these
respective excitations differ by many orders of magnitude, these
approximations are not contradictory. This process gives us 
\begin{eqnarray}
i\hbar\frac{\partial\hat{\rho}}{\partial
  t}&=&\left[\hat{H},\hat{\rho}\right]+\hat{{\cal L}}\hat{\rho}\nonumber\\
&=&\left[\frac{1}{2m}\hat{p}^{2}+\frac{1}{2}m\omega_{m}^{2}\hat{x}^{2}-\hbar
g\hat{a}^{\dag}\hat{a}\hat{x}-i\hbar\epsilon(\hat{a}-\hat{a}^{\dag}),\hat{\rho}\right]\nonumber\\
& &
+i\hbar\gamma\left(2\hat{a}\hat{\rho}\hat{a}^{\dag}-\hat{a}^{\dag}\hat{a}\hat{\rho}
-\hat{\rho}\hat{a}^{\dag}\hat{a}\right)+D_{m}\hat{\rho},
\end{eqnarray}
where $\gamma$ represents the loss rate through the fixed mirror and $D_{m}\hat{\rho}$ represents the
mirror damping, using the Brownian motion master equation developed by Di\'osi~\cite{Diosi},
\begin{equation}
D_{m}\hat{\rho}=\gamma_{m}\left[\hat{x},\{\hat{p},\hat{\rho}\}\right]
-\frac{i\hbar\gamma_{m}}{2\lambda_{dB}^{2}}\left[\hat{x},[\hat{x},\hat{\rho}]\right]
-\frac{i\kappa\gamma_{m}\lambda_{dB}^{2}}{\hbar}\left[\hat{p},[\hat{p},\hat{\rho}]\right].
\label{eq:diosi}
\end{equation}
In the above, $\gamma_{m}$ is the mirror damping rate, which depends on temperature through the mechanical quality factor, $\lambda_{dB}=\hbar/\sqrt{4mk_{B}T}$, the thermal de Broglie wavelength of the mirror, with $k_{B}$ Boltzmann's constant and $T$ the temperature. $\kappa$ is a numerical factor which must be greater than $1$ for this master equation to be of the Lindblad form, but is not important here as it leads to terms of the order $\hbar\omega_{m}/k_{B}T$, which we will show to be insignificant at the temperatures we consider.

\section{Expansion of the mirror in coherent states}
\label{sec:stochastics}

Rather than writing the Heisenberg equations of motion, which are difficult to solve, we will make use of the original definition~\cite{Crispin} of the annihilation and creation operators in terms of $\hat{x}$ and $\hat{p}$ and develop stochastic differential equations in the positive-P representation~\cite{P+}. This allows us to use c-number equations which describe all the quantum properties of the mirror dynamics contained in the original master equation.  
We will describe the operators $\hat{x}$ and $\hat{p}$, in terms of the operators $\hat{b}$ and $\hat{b}^{\dag}$, where
\begin{eqnarray}
\hat{x} &=& A\left(\hat{b}+\hat{b}^{\dag}\right),\nonumber\\
\hat{p} &=& B\left(\hat{b}-\hat{b}^{\dag}\right),
\label{eq:mirrorstates}
\end{eqnarray}
with
\begin{eqnarray}
A &=& \sqrt{\frac{\hbar}{2m\omega_{m}}},\nonumber\\
B &=& -i\sqrt{\frac{\hbar m\omega_{m}}{2}},
\label{eq:A&B}
\end{eqnarray}
and $[\hat{b},\hat{b}^{\dag}]=1$. 
Writing the equations using these variables is advantageous because it allows us to automatically define a P-representation of the density matrix in terms of  an expansion in the minimum uncertainty (coherent) states $|\beta\rangle$, defined as $\hat{b}|\beta\rangle=\beta|\beta\rangle$. It also means that the mirror quadrature variances have a coherent state or vacuum value of $1$, the same as for the electromagnetic field. In fact, $A$ and $|B|$ represent the standard quantum limits (SQL) for measurement of the mirror position and momentum, respectively. We note that these phonon annihilation and creation operators have previously been used to describe the mirror, but not in the context of developing phase-space representation stochastic differential equations~\cite{Bose,macroknight,zhang}.

In terms of these new variables, the master equation for the mirror
damping is now written as
\begin{eqnarray}
D_{m}\hat{\rho} &=& \gamma_{m}AB\left[\hat{b}+\hat{b}^{\dag},\left\{\hat{b}-\hat{b}^{\dag},\hat{\rho}\right\}\right]-\frac{i\hbar\gamma_{m}A^{2}}{2\lambda_{dB}^{2}}\left[\hat{b}+\hat{b}^{\dag},\left[\hat{b}+\hat{b}^{\dag},\hat{\rho}\right]\right]\nonumber\\
& &-\frac{i\kappa\gamma_{m}\lambda_{dB}^{2}}{\hbar}B^{2}[\hat{b}-\hat{b}^{\dag},[\hat{b}-\hat{b}^{\dag},\hat{\rho}]].
\label{eq:betadiosi}
\end{eqnarray}

\section{Stochastic equations}
\label{sec:stochastic}

Using the well-known operator correspondences for the P-representation~\cite{Crispin}, we may map the master equation onto a partial differential equation for the P-function of the system,
\begin{eqnarray}
\frac{dP}{dt} &=& \left\{-\left[\frac{\partial}{\partial\alpha}\left(\epsilon-\gamma\alpha+igA\alpha\left[\beta+\beta^{\ast}\right)\right]
\right.\right.\nonumber\\
& &\left.\left.
+\frac{\partial}{\partial\alpha^{\ast}}\left(\epsilon^{\ast}-\gamma\alpha^{\ast}-igA\alpha^{\ast}\left[\beta+\beta^{\ast}\right]\right)
\right.\right.\nonumber\\
& &\left.\left.
+\frac{\partial}{\partial\beta}\left(-i\omega_{m}\beta+igA|\alpha|^{2}-\gamma_{m}\left[\beta-\beta^{\ast}\right]\right)
\right.\right.\nonumber\\
& &\left.\left.
+\frac{\partial}{\partial\beta^{\ast}}\left(i\omega_{m}\beta^{\ast}-igA|\alpha|^{2}-\gamma_{m}\left[\beta^{\ast}-\beta\right]\right)\right]
\right.\nonumber\\
& &\left.
+\frac{1}{2}\left[\frac{\partial^{2}}{\partial\alpha\partial\beta}\left(-igA\alpha\right)+
\frac{\partial^{2}}{\partial\beta\partial\alpha}\left(-igA\alpha\right)\right.\right.\nonumber\\
& & \left.\left. +
\frac{\partial^{2}}{\partial\alpha^{\ast}\partial\beta^{\ast}}\left(igA\alpha^{\ast}\right)
+\frac{\partial^{2}}{\partial\beta^{\ast}\partial\alpha^{\ast}}\left(igA\alpha^{\ast}\right)
\right.\right.\nonumber\\
& &\left.\left.
+\frac{\partial^{2}}{\partial\beta^{2}}\left(\gamma_{m}\left[1-\frac{2k_{B}T}{\hbar\omega_{m}}+\frac{\kappa\hbar\omega_{m}}{4k_{B}T}\right]\right)
+\frac{\partial^{2}}{\partial\beta^{\ast 2}}\left(\gamma_{m}\left[1-\frac{2k_{B}T}{\hbar\omega_{m}}+\frac{\kappa\hbar\omega_{m}}{4k_{B}T}\right]\right)
\right.\right.\nonumber\\
& &\left.\left.
+\frac{\partial^{2}}{\partial\beta\partial\beta^{\ast}}\left(-\gamma_{m}\left[1-\frac{2k_{B}T}{\hbar\omega_{m}}-\frac{\kappa\hbar\omega_{m}}{4k_{B}T}\right]\right)\right.\right.\nonumber\\
& &\left.\left.
+\frac{\partial^{2}}{\partial\beta^{\ast}\partial\beta}\left(-\gamma_{m}\left[1-\frac{2k_{B}T}{\hbar\omega_{m}}-\frac{\kappa\hbar\omega_{m}}{4k_{B}T}\right]\right)
\right]\right\}P(\alpha,\alpha^{\ast},\beta,\beta^{\ast},t).
\label{eq:betaplanck}
\end{eqnarray}
The diffusion matrix of the above equation is
\begin{equation}
D=\left[
\begin{array}{cccc}
0 &0 &-igA\alpha &0\\
0 &0 &0 &igA\alpha^{\ast}\\
-igA\alpha &0 &\gamma_{m}\left(1-\frac{2k_{B}T}{\hbar\omega_{m}}+\frac{\kappa\hbar\omega_{m}}{4k_{B}T}\right) & -\gamma_{m}\left(1-\frac{2k_{B}T}{\hbar\omega_{m}}-\frac{\kappa\hbar\omega_{m}}{4k_{B}T}\right)
\\
0 &igA\alpha^{\ast} & -\gamma_{m}\left(1-\frac{2k_{B}T}{\hbar\omega_{m}}-\frac{\kappa\hbar\omega_{m}}{4k_{B}T}\right)
&\gamma_{m}\left(1-\frac{2k_{B}T}{\hbar\omega_{m}}+\frac{\kappa\hbar\omega_{m}}{4k_{B}T}\right)
\end{array}
\right],
\label{eq:betadiff}
\end{equation}
We note here that this drift matrix has diverging terms as $T\rightarrow 0$, but this is not a problem as the Di\'{o}si master equation is valid in the limit where $k_{B}T\gg\hbar\omega_{m}$.
As a physical example, in Ref.~\cite{Schiller}, we find $\omega_{m}=1.6\times10^{5}$\;s$^{-1}$, so that $\hbar\omega_{m}=1.72\times10^{-29}$\;J, whereas $k_{B}T=5.8\times10^{-23}$\;J at $4.2$\;K, the temperature which we will mainly use in our investigations.

If we wish to treat Eq.~\ref{eq:betaplanck} as a genuine Fokker-Planck equation which we may map onto stochastic differential equations, the matrix $D$ must be positive-definite.
Numerical investigations using typical parameters show that this is not the case,
therefore
for quantum calculations we will have to use the positive-P representation~\cite{P+}.
The positive-P representation equations in a doubled phase space can be found by the simple change of variables $\alpha^{\ast}\rightarrow\alpha^{+},\:\beta^{\ast}\rightarrow\beta^{+}$, so that, (noting that $\overline{\alpha^{+}}=\overline{\alpha^{\ast}}$ only in the mean and similarly for $\beta^{+}$), we now have four independent stochastic variables. Ignoring the terms proportional to $\hbar\omega_{m}/k_{B}T$ due to their small relative size, one possible factorisation of the diffusion matrix, $D=NN^{\mbox{T}}$, of Eq.~\ref{eq:betadiff} is
\begin{equation}
N=\left[
\begin{array}{ccccc}
0 &\sqrt{\frac{-igA\alpha}{2}} &\sqrt{\frac{igA\alpha}{2}} &0 &0\\
0 &0 &0 &\sqrt{\frac{igA\alpha^{+}}{2}} &-\sqrt{\frac{-igA\alpha^{+}}{2}}\\
-\sqrt{\gamma_{m}\left(1-\frac{2k_{B}T}{\hbar\omega_{m}}\right)} &\sqrt{\frac{-igA\alpha}{2}} &-\sqrt{\frac{igA\alpha}{2}} &0 &0\\
\sqrt{\gamma_{m}\left(1-\frac{2k_{B}T}{\hbar\omega_{m}}\right)} &0 &0 &\sqrt{\frac{igA\alpha^{+}}{2}} &\sqrt{\frac{-igA\alpha^{+}}{2}}
\end{array}
\right],
\label{eq:noisemat}
\end{equation}
which allows us to write a set of four stochastic differential
equations (Note that the It\^o form and the Stratonovich form of these
equations are identical.),
\begin{eqnarray}
\frac{d\alpha}{dt} &=& \epsilon-\gamma\alpha+igA\alpha(\beta+\beta^{+})+\sqrt{\frac{-igA\alpha}{2}}(\eta_{2}+i\eta_{3}),
\nonumber\\
\frac{d\alpha^{+}}{dt} &=& \epsilon^{\ast}-\gamma\alpha^{+}-igA\alpha^{+}(\beta+\beta^{+})+\sqrt{\frac{igA\alpha^{+}}{2}}(\eta_{4}-i\eta_{5}),\nonumber\\
\frac{d\beta}{dt} &=& -i\omega_{m}\beta-\gamma_{m}(\beta-\beta^{+})+igA\alpha^{+}\alpha\nonumber\\
& &
-\sqrt{\gamma_{m}\left(1-\frac{2k_{B}T}{\hbar\omega_{m}}\right)}\;\eta_{1}+\sqrt{\frac{-igA\alpha}{2}}(\eta_{2}-i\eta_{3}),\nonumber\\
\frac{d\beta^{+}}{dt} &=& i\omega_{m}\beta^{+}+\gamma_{m}(\beta-\beta^{+})-igA\alpha^{+}\alpha\nonumber\\
& &
+
\sqrt{\gamma_{m}\left(1-\frac{2k_{B}T}{\hbar\omega_{m}}\right)}\;\eta_{1}+\sqrt{\frac{igA\alpha^{+}}{2}}(\eta_{4}+i\eta_{5}).
\label{eq:PPSDE}
\end{eqnarray}
In the above, the real Gaussian noise terms have the correlations
\begin{eqnarray}
\overline{\eta_{i}(t)} &=& 0,\nonumber\\
\overline{\eta_{i}(t)\eta_{j}(t')} &=& \delta_{ij}\delta(t-t').
\label{eq:noisedef}
\end{eqnarray}
The set of coupled equations (\ref{eq:PPSDE}) may be integrated numerically, with averages taken over a large number of stochastic trajectories, which allows for the probabilistic calculation of any desired normally-ordered operator moments. As an example, with N trajectories, we have
\begin{equation}
\langle\hat{a}^{\dag\;m}\hat{a}^{n}\rangle=\lim_{N\rightarrow\infty}\frac{1}{N}\sum_{j=1}^{N}\alpha^{+\;m}\alpha^{n},
\label{eq:averaging}
\end{equation}
where $j$ labels the results from the $j$th trajectory. 

\section{Classical analysis}
\label{sec:classic}

\subsection{Steady-state solutions}
\label{sec:ssclassic}

Before we return to the full stochastic equations, which we will solve numerically, we will investigate some of the classical properties of the system, which allow for analytical insights.
From the drift part of Eq.~\ref{eq:betaplanck}, we can immediately write the mean-field equations using the notation $\overline{z}$ for the classical mean-field value of $z$,
\begin{eqnarray}
\frac{d\overline{\alpha}}{dt} &=& \epsilon-\gamma\overline{\alpha}+igA\overline{\alpha}(\overline{\beta}+\overline{\beta^{\ast}}),
\nonumber\\
\frac{d\overline{\beta}}{dt} &=& -i\omega_{m}\overline{\beta}-\gamma_{m}(\overline{\beta}-\overline{\beta^{\ast}})+igA\overline{|\alpha|^{2}},
\label{eq:meanbeta}
\end{eqnarray}
from which we may find the classical steady state solutions.

Solving Eqs.~\ref{eq:meanbeta} for the steady-states, we find that $\beta_{ss}$ is real, which means that the steady-state momentum is zero. (Note that this will not be the prediction of stochastic integration of the full equations.) However, using this fact we may write the solutions as
\begin{eqnarray}
\beta_{ss} &=& \beta_{ss}^{\ast} = \frac{gA}{\omega_{m}}|\alpha_{ss}|^{2},\nonumber\\
\alpha_{ss} &=& \frac{\epsilon}{\gamma-2igA\beta_{ss}}.
\label{eq:ssbeta}
\end{eqnarray}
Although the solutions above are not closed (the solution for $\alpha_{ss}$ is a function of $\beta_{ss}$, etc.), we can make an iterative expansion, beginning with the result for fixed mirrors,
\begin{equation}
\alpha_{ss}^{0}=\frac{\epsilon}{\gamma},
\label{eq:alpha0}
\end{equation}
and substitute this into the solution for $\beta_{ss}$. This can then be substituted into the solution for $\alpha_{ss}$, the process being repeated until we attain the required degree of convergence. We note here that there are parameter regimes for which this expansion does not converge and that these are regions where we do not find classical steady-state solutions, but rather a limit-cycle, self-pulsing behaviour~\cite{Claude}. 

Although we will demonstrate below that the classical steady-state solutions, especially for $x$ and $p$, are not accurate in any parameter regime, they do allow for some insight into which property of the electromagnetic field outside the cavity is most likely to allow for an inference of the mirror position. The usual candidates are the intensity, which may be measured by photodetection, and the quadratures, which may be measured by homodyne detection. Defining the intracavity quadratures as $\hat{X}_{a}=\hat{a}+\hat{a}^{\dag}$ and $\hat{Y}_{a}=-i(\hat{a}-\hat{a}^{\dag})$, with their classical equivalents written in terms of $\alpha_{ss}$ and $\alpha_{ss}^{\ast}$, we find
\begin{eqnarray}
X_{a} &=& \frac{2\epsilon\gamma}{\gamma^{2}+g^{2}x^{2}},\nonumber\\
Y_{a} &=& \frac{2g\epsilon x}{\gamma^{2}+g^{2}x^{2}},\nonumber\\
|\alpha|^{2} &=& \frac{\epsilon^{2}}{\gamma^{2}+g^{2}x^{2}},
\label{eq:XYN}
\end{eqnarray}
where we have set $\epsilon$ as real and used $x=2A\beta$, with $\beta$ real. As $\gamma^{2}$ is typically much larger than $g^{2}x^{2}$, we may make a series expansion of these expressions. We find that
\begin{eqnarray}
X_{a} &\approx& \frac{2\epsilon}{\gamma}\left(1-\frac{g^{2}x^{2}}{\gamma^{2}}\right),\nonumber\\
Y_{a} &\approx& \frac{2g\epsilon x}{\gamma^{2}}\left(1-\frac{g^{2}x^{2}}{\gamma^{2}}\right),\nonumber\\
|\alpha|^{2} &\approx& \frac{\epsilon^{2}}{\gamma^{2}}\left(1-\frac{g^{2}x^{2}}{\gamma^{2}}\right).
\label{eq:expansion}
\end{eqnarray}
It is immediately obvious that $Y_{a}$ depends on $x$ to first order, while the other two exhibit only a second order dependence. This shows that homodyne measurements of the $Y_{a}$ quadrature will be more sensitive to variations in the position of the mirror than will be the other two measurements, as previously noted by Vitali {\em et al.\/}~\cite{Ribichini}, although the $X_{a}$ quadrature will show a weaker dependence rather than being totally independent of $x$ as in the linearised analysis of Ref.~\cite{Ribichini}.
  
\subsection{Bistability}
\label{eq:biestavel}

It has been predicted that, with a non-zero detuning between the field and the cavity resonance, this system can exhibit bistability in the optical intensity~\cite{nariz}. 
To find the condition for bistability, we start with the classical
equations with detuning, $\Delta$, included,
\begin{eqnarray}
\frac{d\overline{\alpha}}{dt} &=& \epsilon-(\gamma+i\Delta)\overline{\alpha}+igA\overline{\alpha}(\overline{\beta}+\overline{\beta^{\ast}}),
\nonumber\\
\frac{d\overline{\beta}}{dt} &=& -i\omega_{m}\overline{\beta}-\gamma_{m}(\overline{\beta}-\overline{\beta^{\ast}})+igA\overline{|\alpha|^{2}}.
\label{eq:detunebeta}
\end{eqnarray}
These have the steady-state solutions,
\begin{eqnarray}
\beta_{ss}&=&\beta_{ss}^{\ast}=\frac{gA}{\omega_{m}}|\alpha_{ss}|^{2},\nonumber\\
|\alpha_{ss}|^{2}&=&\frac{\epsilon^{2}}{\gamma^{2}+\left(\Delta-2gA\beta_{ss}\right)^{2}},
\label{eq:estacionario}
\end{eqnarray} 
which immediately leads to the cubic equation in $I=|\alpha_{ss}|^{2}$,
\begin{equation}
\frac{4g^{4}A^{4}}{\omega_{m}^{2}}I^{3}-\frac{4g^{2}A^{2}\Delta}{\omega_{m}}I^{2}+(\gamma^{2}+\Delta^{2})I-\epsilon^{2}=0.
\label{eq:cubic}
\end{equation}
We find the condition for bistability by differentiating this expression with respect to $I$, which gives
\begin{equation}
\frac{12g^{4}A^{4}}{\omega_{m}^{2}}I^{2}-\frac{8g^{2}A^{2}\Delta}{\omega_{m}}I+(\gamma^{2}+\Delta^{2})=0.
\label{eq:Idiff}
\end{equation}
The condition for bistability is that this quadratic equation has two positive real roots. The roots are written as
\begin{equation}
r_{\pm}=\frac{\Delta\omega_{m}}{3g^{2}A^{2}}\pm\frac{\omega_{m}}{6g^{2}A^{2}}\sqrt{\Delta^{2}-3\gamma^{2}},
\label{eq:raizes}
\end{equation}
giving the inequality
\begin{equation}
\Delta\pm\frac{1}{2}\sqrt{\Delta^{2}-3\gamma^{2}}>0.
\label{eq:desigualdade}
\end{equation}
A necessary, but not sufficient condition is that $\Delta$ be positive, as $I$ must be positive. This immediately contradicts the condition given by Mancini and Tombesi~\cite{nariz}, $|\Delta|>\sqrt{3}\gamma$, which allows for negative intensities. As $I$ must also be real, we find the condition for bistability as $\Delta>\sqrt{3}\gamma$. 

\section{Stochastic results}
\label{sec:estocastica}

\subsection{Initial conditions}
\label{sec:inicial}

To numerically integrate Eq.~\ref{eq:PPSDE}, we make use of the fact
that the It\^o and Stratonovich forms are identical so that we may use a standard three-step
predictor-corrector method. The convergence of the algorithm was
checked by comparison with a four-step method, and also by halving the
time-step. In all the quantities shown, the sampling errors are
comparable to 
the thickness of the plotted lines.
 
We will use the published experimental parameters of Ref.~\cite{Schiller}, and make comparisons with the theoretical predictions reported elsewhere. The oscillating mirror is considered as being perfectly reflecting, with a mass of $m=10^{-5}$\;kg, a mechanical quality factor of $Q=4\times10^{6}$ at $4.2$\;K, decreasing to $2.25\times10^{6}$ at $70$\;K, and a resonance frequency $\omega_{m}/2\pi=26$\;kHz. The damping rate of the mirror is $\gamma_{m}=.5\omega_{m}/Q=0.0363$\;s$^{-1}$ at $4.2$\;K. We consider a cavity length of $L=1$\;cm, with cavity finesse ${\cal F}=15\times10^{3}$, which gives $\gamma=\pi c/2{\cal F}L=3.14\times10^{6}$\;s$^{-1}$. We consider an optical wavelength of $\lambda=1064$\;nm, which gives $\omega_{0}=1.77\times10^{15}$\;s$^{-1}$, and a coupling between the light and the mirror of $g=\omega_{0}/L=1.77\times10^{17}$\;m$^{-1}$s$^{-1}$. The optical pumping of the cavity is $\epsilon=\sqrt{\gamma{\cal P}/\hbar\omega_{0}}$, where ${\cal P}$ is the laser power in Watts.

In stochastic integration of the equations which describe an intracavity optical system, the standard approach is to begin with the state inside the cavity as vacuum so that, with a continuous pump, the system enters the steady-state (or its limit cycle behaviour in the case of self-pulsing) after a few cavity lifetimes. In the present case the situation is somewhat different, as not only the electromagnetic field, but also the oscillating mirror, has to reach the steady-state. As the relaxation time of the mirror can be orders of magnitude larger than that of the intracavity field, and we need to average over a large number of trajectories to obtain reliable results, it is not practical to begin the integration with an arbitrary initial condition for the mirror. Naively beginning with $\beta(0)=\beta^{+}(0)=0$, the ground state of the mirror, leads to extremely long lived transients, as this is far from the equilibrium state at finite temperature. To give some idea, even at $4.2$\;K, which is perhaps the lowest easily achievable temperature, the average number of mirror quanta becomes $\overline{|\beta|}^{2}=k_{B}T/\hbar\omega_{m}= 3.36\times10^{6}$. For purposes of comparison, we will therefore choose the initial mirror state in two different ways and integrate the equations without any pumping of the cavity. Firstly, as a (real) coherent state, which has uncertainties in position and momentum at the SQL, with the P-function
\begin{equation}
P(\beta)=\delta(\beta-\sqrt{k_{B}T/\hbar\omega_{m}}),
\label{eq:coherent}
\end{equation} 
and, secondly from the thermal distribution
\begin{equation}
P(\beta)=\frac{1}{\pi\overline{n}}\mbox{e}^{-|\beta|^{2}/\overline{n}},
\label{eq:Pthermal}
\end{equation}
where $\overline{n}=k_{B}T/\hbar\omega_{m}$. Note that, at the beginning of each trajectory, $\beta=(\beta^{+})^{\ast}$ and the phase is completely random for the thermal distribution.
We stress here that the variance in the number of mirror phonons for a thermal state is $V(n)=\overline{n}^{2}+\overline{n}$, which is very much larger than $\overline{n}$. Using the Planck distribution,
\begin{equation}
\overline{n}=\left(\mbox{e}^{\hbar\omega_{m}/k_{B}T}-1\right)^{-1},
\label{eq:Max}
\end{equation}
we find that to achieve $\overline{n}=1$, we would need $T=1.8\;\mu$K, and even then the variance would be $2$, or twice the mean value.

It is important to note here that the number of phonons does not enter into the equations, but rather the quadratures $\hat{X}_{b}$ and $\hat{Y}_{b}$. In a linearised approach using our equations, it is the uncertainties in these which are important. We can easily calculate these for a thermal state of the mirror with an unpumped cavity. A simple integration gives
\begin{equation}
V(\hat{X}_{b})=V(\hat{Y}_{b})=1+2\sqrt{\pi}\left(\frac{k_{B}T}{\hbar\omega_{m}}\right)^{3/2},
\label{eq:VXYbthermal}
\end{equation}
equal to $2.2\times10^{10}$ at $T=4.2$\;K for our system. This is in stark contrast to a coherent state of the mirror, sometimes used to facilitate the mathematics of a linearised analysis, and for which $V(\hat{X}_{b})=V(\hat{Y}_{b})=1$.

\begin{figure}
\begin{center} 
\includegraphics[width=0.8\columnwidth]{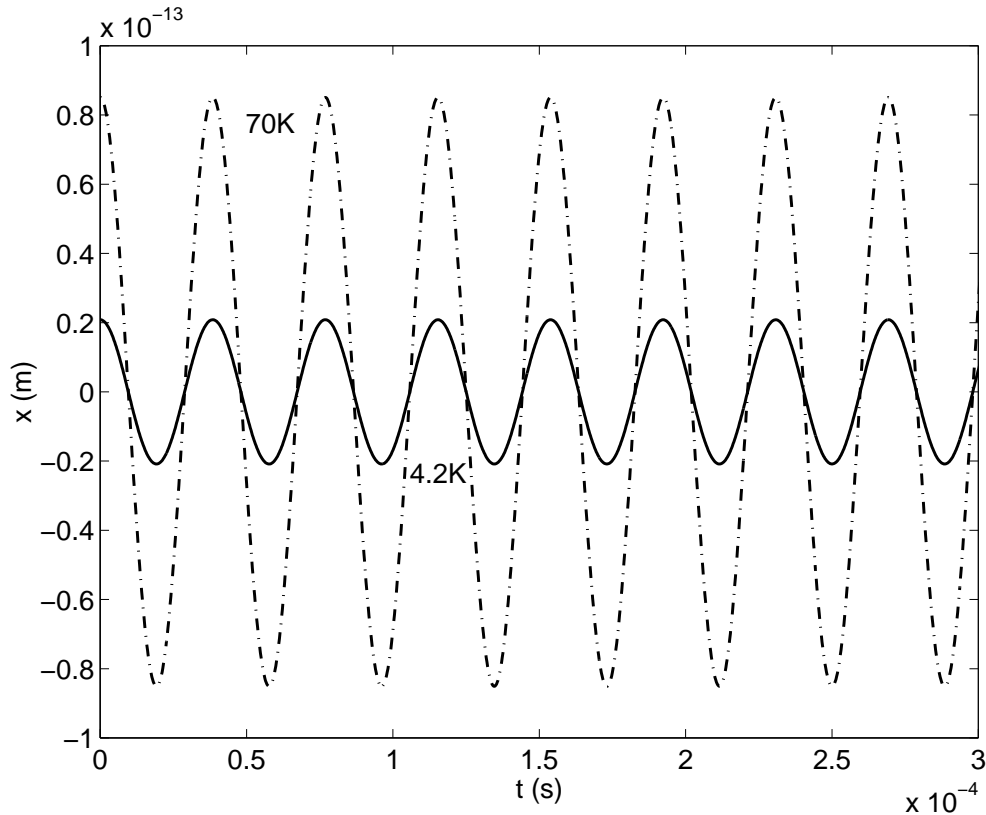}
\end{center}  
\caption{Mean values for $x$ at $T=4.2$\;K and $T=70$\;K for an
  initial coherent state of the mirror and no optical pumping. These
  results are the averages of $7.6\times10^{3}$ and $2.7\times10^{4}$
  trajectories, respectively. Note that, unless otherwise stated, the values plotted here and in subsequent graphs are dimensionless.}
\label{fig:xpumpco0}
\end{figure} 

\begin{figure}
\begin{center} 
\includegraphics[width=0.8\columnwidth]{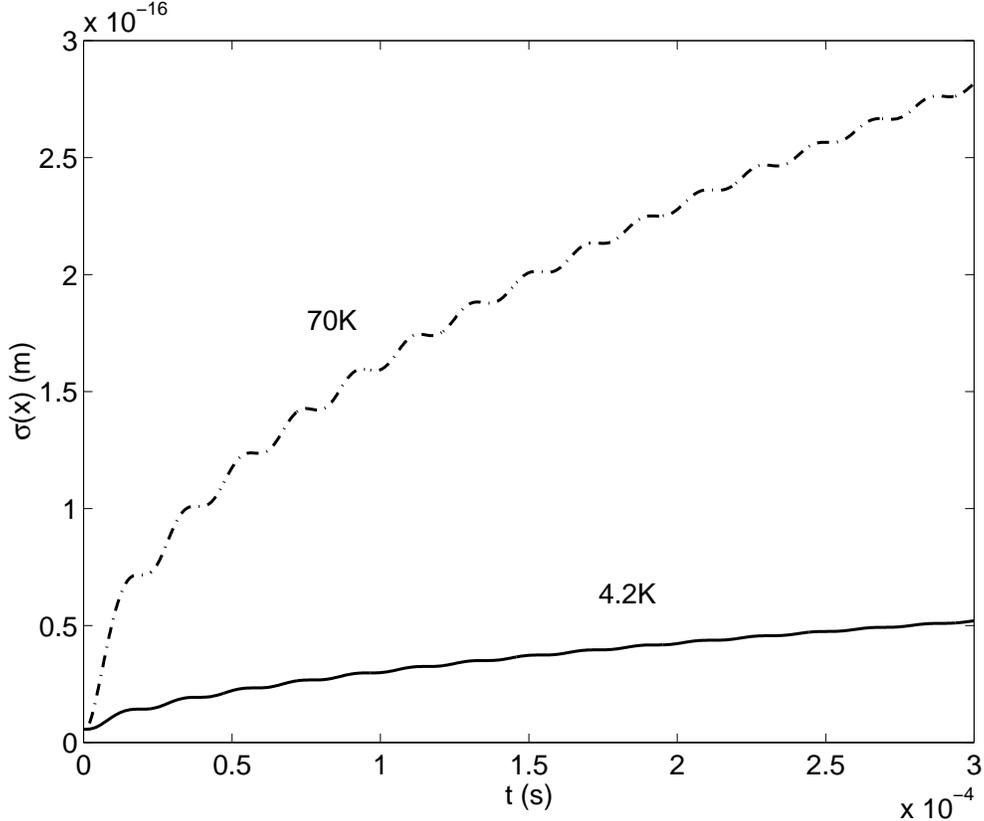}
\end{center}  
\caption{Standard deviations for $x$ at $T=4.2$\;K and $T=70$\;K for
  an initial coherent state of the mirror and no optical pumping. Note
  that these quantities were still increasing at twice the time shown
  here and would be expected to eventually attain the thermal values.}
\label{fig:Vxpumpco0}
\end{figure}

We have calculated the stochastic results for the means and standard
deviations of the mirror position, without any optical pumping and for
initial thermal and coherent states of the mirror at temperatures of $4.2$\;K and $70$\;K.  
In Fig.~\ref{fig:xpumpco0} we show the
stochastic results for the position, $x$, of the mirror, with
initial coherent states at these two temperatures. Considering only
these mean values could give the erroneous impression that the mirror
is 
in a non-stationary steady-state, which we can immediately see is not
the case when we look at Fig.~\ref{fig:Vxpumpco0}, which shows the
standard deviations, $\sigma(x)$, for the same parameters. Although we
have shown the standard deviations here, the variances
for the initial coherent state continue to increase linearly for more
than twice the time shown, which was as far as we continued the
integration. In contrast, for an initial thermal state, the mean value
of the mirror displacement is, by definition, zero as can be seen from the equation
for the P-function (\ref{eq:Pthermal}). We note here that our stochastic results over
more than $2\times 10^{6}$ trajectories still showed oscillations of the order of
$10^{-16}$\;m, but that we are confident that this small, but non-zero, value is
due to the difficulty of sampling the distribution with a finite
number of trajectories. By comparison with the coherent state values,
for an initial thermal state 
$\sigma(x)\approx 1\times10^{-14}$\;m at $T=4.2$\;K, and is almost constant, indicating that this is a good choice of initial condition. This value agrees well with the expression given in Ref.~\cite{Schiller} for the thermal noise, $\sigma(x)=\sqrt{k_{B}T/m\omega_{m}^{2}}$, which gives a value of $1.47\times10^{-14}$\;m. Note that, over the time scales shown in Fig.~\ref{fig:xpumpco0} we do not see any decay in the oscillations towards the thermal state values, as this would be expected to happen on a time scale of $1/\gamma_{m}$, which is approximately $50$\;s for the parameters used here. In fact, although an initial coherent state of the mirror has been used in theoretical analyses (see, for example, Bose {\em et al.\/}~\cite{Bose}), it is not at all obvious how this particular state may be constructed experimentally. A thermal state arises naturally, and will be equal to a coherent state for $T=0$\;K, but absolute zero cannot be reached experimentally. In optics, a coherent state can be described theoretically as a displacement of the vacuum by the displacement operator $D(\alpha)=\exp(\alpha\hat{a}^{\dag}-\alpha^{\ast}\hat{a})$, which we can see has some relation to the optical pumping term of the Hamiltonian, $H_{pump}$. Therefore an ideal empty cavity with this pumping term will naturally develop an intracavity coherent state. We are not aware of any similar candidate for the mirror, even if it could begin in the $T=0$ vacuum state. Therefore we will use an initial thermal state in our investigations.

\begin{figure}
\begin{center} 
\includegraphics[width=0.8\columnwidth]{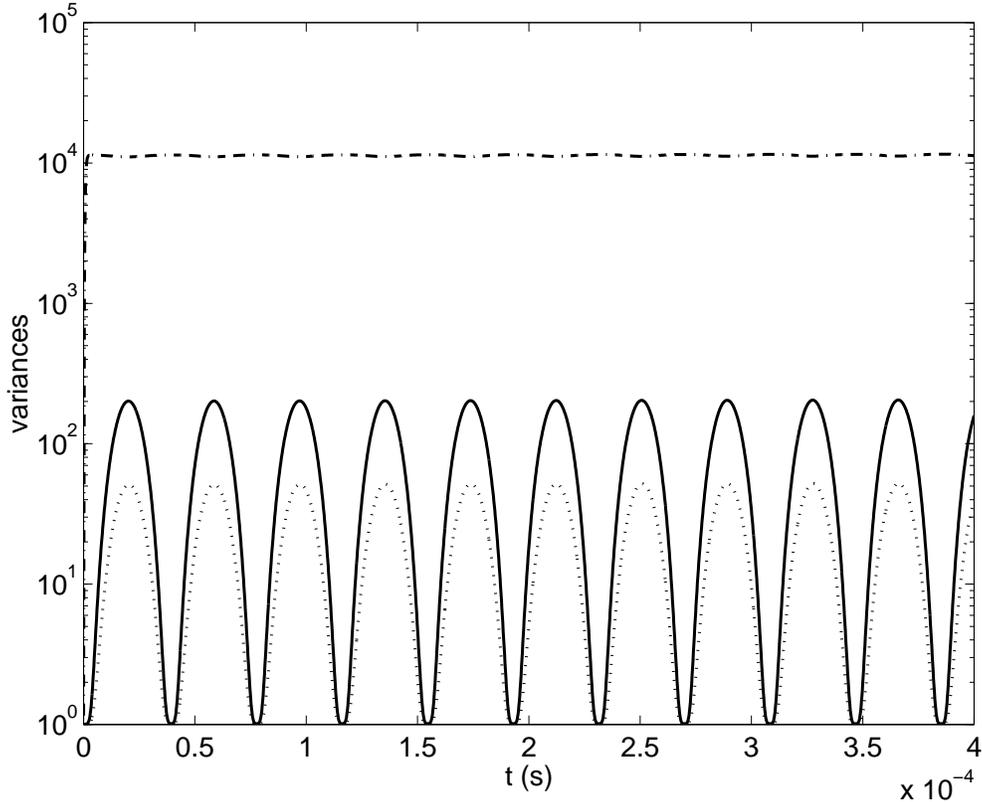}
\end{center}  
\caption{The variances of the intracavity electromagnetic field for a laser power of $5$\;mW. The solid line is $V(\hat{X}_{a})$, the dash-dotted line is $V(\hat{Y}_{a})$, and the dotted line is the Fano factor.}
\label{fig:luzVar}
\end{figure}

When we examine the  stochastic results for the intracavity field intensity, for an input power ${\cal P}=5$\;mW, we find that the field exhibits a self-pulsing behaviour at approximately the resonance frequency of the mirror, as previously predicted~\cite{Claude}. However, the oscillations are of very small relative amplitude, at approximately $0.2\%$ of the average mean intensity. With increasing input power, the oscillations become larger until, for a power of $100$mW, for example, they are more than half the maximum intensity. At the lower power, the mean motion of the mirror is an oscillation between $0$ and $1.2\times10^{-12}$m, while at the higher power oscillates between $-1$ and $3\times10^{-11}$m. Even though these displacements are truly microscopic, they have a noticeable effect on the mean intensity, which should be easily detectable experimentally. Interestingly enough, these results are almost identical to what we find by numerical integration of the classical equations (\ref{eq:meanbeta}), although these can tell us nothing about the quantum correlations which we wish to investigate. Among these quantum correlations are the variances of the intracavity field and the Fano factor, defined as $F(N_{a})=V(N_{a})/N_{a}$. These results, which we averaged over $6.71\times10^{5}$ trajectories,  are 
shown in Fig.~\ref{fig:luzVar}. For a coherent state, all three values are $1$, which would be zero on the logarithmic vertical scale used here. As all three correlations are greater than or equal to one, we do not see any squeezing of the field in the time domain, but do see excess noise in all three quantities. As is common with Kerr media, there is more excess noise in the $\hat{Y}_{a}$ quadrature than in either the $\hat{X}_{a}$ quadrature or in the intensity.

\subsection{Position measurements}
\label{sec:position}

Much of the theoretical and experimental interest in this system has been in indirectly measuring small displacements or small forces which act on the oscillating mirror, by means of measurements on the optical field. Generally theoretical results are presented in terms of output spectra which allow for an inferred value of the mirror position at various frequencies. These spectra are simple to calculate in a linearised analysis which treats the system as an Ornstein-Uhlenbeck process~\cite{Danbook}, but we do not consider linearisation valid here, for the reasons we have stated. To calculate spectra from the results of stochastic integration is also possible in many cases, but here is made difficult by the stiffness of our equations, where the field and the mirror oscillate on vastly different time scales. The length of time needed to integrate a large enough number of trajectories over a sufficient time to give reliable results upon Fourier transformation is prohibitive. Hence we will show results which were obtained in the time domain. 

\begin{figure}
\begin{center} 
\includegraphics[width=0.8\columnwidth]{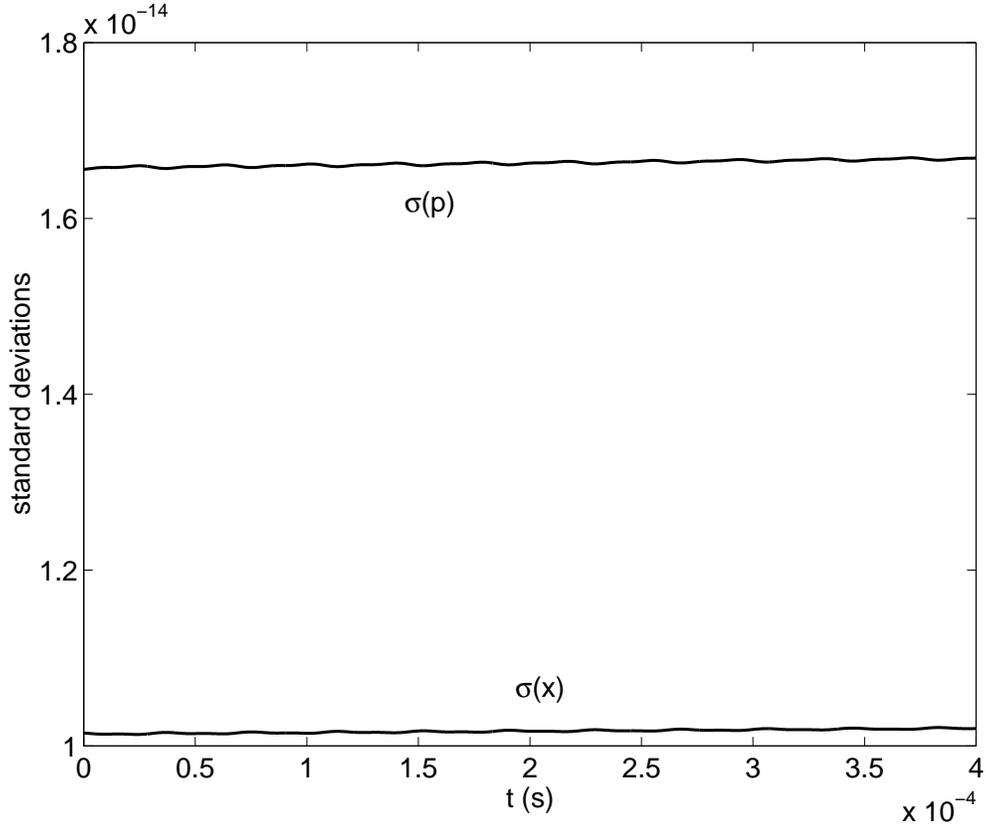}
\end{center}  
\caption{The standard deviations in the mirror position and momentum for a laser power of $5$\;mW. The standard S.I. units for position and momentum are used. Note that a minimum uncertainty state of the mirror has $\sigma(x)=5.6824\times10^{-18}$\;m and $\sigma(p)=9.283\times10^{-18}$\;kgms$^{-1}$.}
\label{fig:sigxp}
\end{figure}

The first results we show for the mirror, in Fig.~\ref{fig:sigxp}, are for the uncertainties in the mirror position and momentum, defined as $\sigma(x)=A\sqrt{V(X_{b})}$ and $\sigma(p)=|B|\sqrt{V(Y_{b})}$. The interaction with the field has not noticeably changed these quantities from the values we found via stochastic integration in the unpumped thermal state case, but has given the momentum a mean value which oscillates between approximately $\pm 1\times10^{-19}$\;kgms$^{-1}$, whereas it was stable at zero in the unpumped case. It is readily seen that the fluctuations in momentum are several orders of magnitude larger than the mean value. We also note here that the uncertainties shown are certainly underestimated, due to the difficulties of simulating the large $\beta$ values with small probabilities in the initial conditions used. This result also shows that, when the likely thermal state of the mirror is taken into account, the back-action noise of the field on the mirror has very little effect at the laser power considered here. At an increased power of $100$\;mW the standard deviations in both position and momentum both increase due to this back-action noise, reaching oscillatory values around means of $\approx1\times10^{-13}$, while the mean momentum oscillates between $\pm3\times10^{-17}$\;kgms$^{-1}$.      

\begin{figure}
\begin{center} 
\includegraphics[width=0.8\columnwidth]{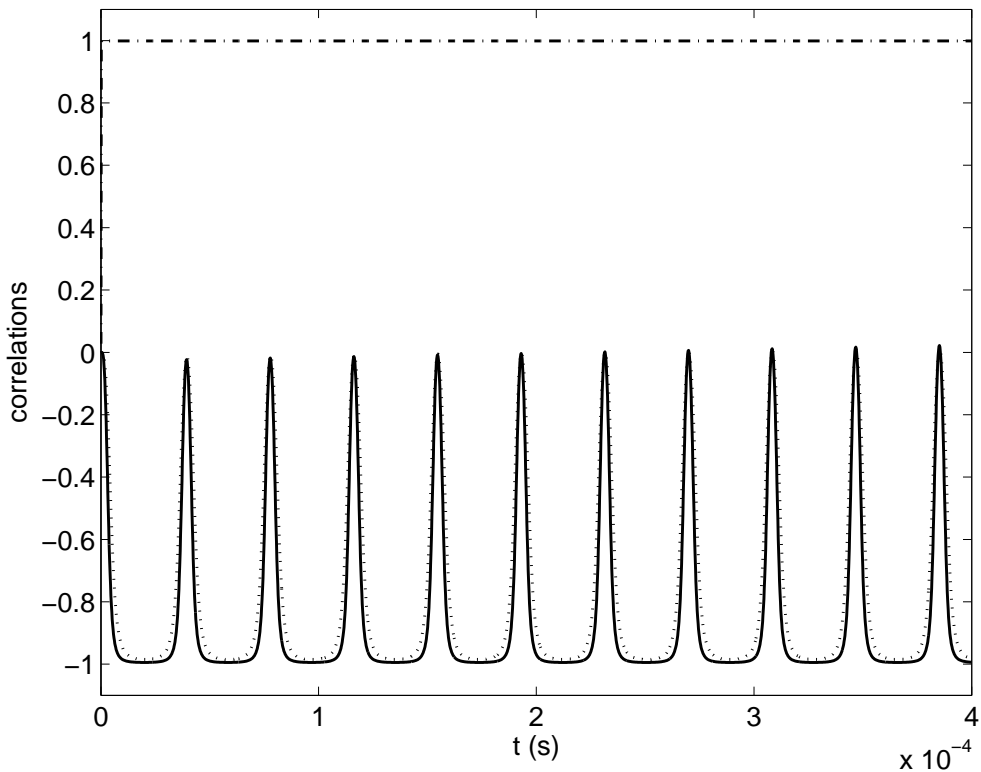}
\end{center}  
\caption{The intracavity correlation functions, ${\cal C}(xX_{a})$ (solid line), ${\cal C}(xY_{a})$ (dash-dotted line), and ${\cal C}(xN_{a})$ (dashed line), for an input power of $5$\;mW.}
\label{fig:corrfun}
\end{figure}

We next investigate the degree of correlation between the position of the mirror, the intracavity optical intensity, and the $\hat{X}_{a}$ and $\hat{Y}_{a}$ quadratures. This correlation function between two quantities $w$ and $z$ is defined as
\begin{equation}
{\cal C}(wz)=\frac{\langle wz\rangle-\langle w\rangle\langle z\rangle}{\sqrt{V(w)V(z)}},
\label{eq:correlationfunction}
\end{equation}
where a perfect correlation gives a value of $1$, a perfect anticorrelation gives a value of 
$-1$ and zero signifies no correlation at all. In Fig.~\ref{fig:corrfun} we see that, as predicted by Eq.~\ref{eq:expansion}, the strongest correlation is between $\hat{Y}_{a}$ and the mirror position, these two being almost perfectly correlated. The correlation functions ${\cal C}(xX_{a})$ and  ${\cal C}(xN_{a})$ oscillate at the frequency $\omega_{m}$ between a value of almost minus one and a value of approximately zero, ranging from being almost perfectly anticorrelated to almost perfectly uncorrelated. This behaviour is quite different from that shown in a linearised fluctuation analysis, where we found that all three of these functions showed anticorrelation around the mirror frequency and were zero at other frequencies. The difference between our fully quantum nonlinear results and the usual linearised predictions is dramatically demonstrated when we consider $|{\cal C}(X_{a}Y_{b})|^{2}$, used to analyse a possible quantum nondemolition measurement by Jacobs {\em et al.\/}~\cite{Kurt}. Their prediction (see their Eq. 31) is for a maximum spectral value of 
$0.9757$ for the parameters we use in this work, whereas our stochastic prediction in the time domain oscillates between $0$ and $6\times10^{-3}$, showing almost no correlation at all. We think that it is unlikely that this difference can be explained by the difference between a spectral measurement and a time domain measurement, but that it is due to the inappropriateness of the linearisation procedure for this system.  

\begin{figure}
\begin{center} 
\includegraphics[width=0.8\columnwidth]{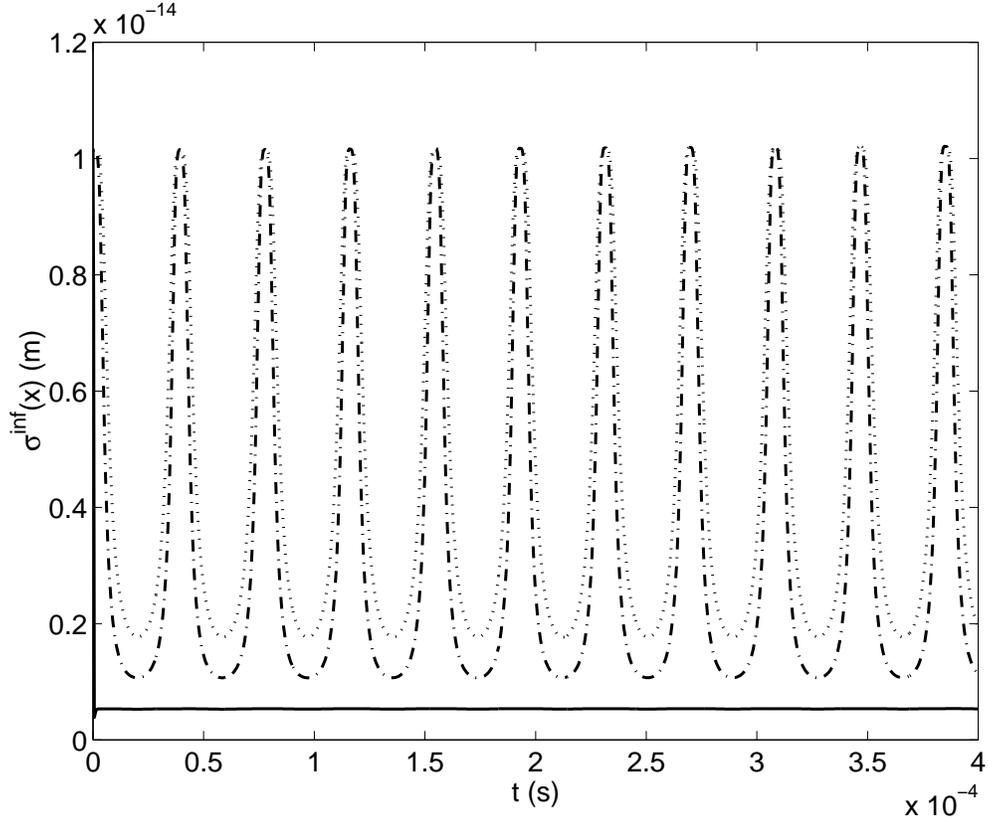}
\end{center}  
\caption{The inferred uncertainties in the mirror position, $A\sqrt{V^{inf}(\hat{X}_{b})}$, calculated as in Eq.~\ref{eq:VXinf}. The solid line is the estimate using $\hat{Y_{a}}$, the dash-dotted line uses $\hat{X}_{a}$ and the dotted line uses $N_{a}$.}
\label{fig:xinfer}
\end{figure}

Another way to infer the mirror position is by linear estimation following measurements of the optical field. We follow a method proposed by Reid~\cite{MDR} in the context of a demonstration of the Einstein-Podolsky-Rosen paradox, and also outlined in Dechoum 
{\em et al.\/}~\cite{turcosafado}. We assume that a measurement of the $\hat{Y}_{a}$ quadrature allows for a linear estimate of $\hat{X}_{b}$, $\hat{X}_{b}^{est}=c\hat{Y}_{a}+d$. This is consistent with the expansion of $Y_{a}$ given above, in Eq.~\ref{eq:expansion}. After optimising for $d$, the RMS error in this estimate is given by 
\begin{equation}
V^{inf}(\hat{X}_{b})=\langle(\hat{X}_{b}-c\hat{Y}_{a})^{2}\rangle-\langle\hat{X}_{b}-c\hat{Y}_{a}\rangle^{2},
\label{eq:RMS}
\end{equation}
which we may minimise as a function of $c$, finding
\begin{equation}
c=\frac{V(\hat{X}_{b},\hat{Y}_{a})}{V(\hat{Y}_{a})}.
\label{eq:choosec}
\end{equation}
We may then write
\begin{equation}
V^{inf}(\hat{X}_{b})=V(\hat{X}_{b})-\frac{[V(\hat{X}_{b},\hat{Y}_{a})]^{2}}{V(\hat{Y}_{a})},
\label{eq:VXinf}
\end{equation}
a quantity which we may calculate via stochastic integration. The inferred uncertainty in a measurement of $\hat{x}$ will then be $\sigma^{inf}(x)=A\sqrt{V^{inf}(\hat{X}_{b})}$, which will be equal to the standard quantum limit when the mirror is inferred to be in a minimum uncertainty state. We have also calculated the inferred position uncertainty using measurements of $\hat{X}_{a}$ and $N_{a}$ in the same manner. As shown in Fig.~\ref{fig:xinfer}, and expected from the previous discussion, using the $\hat{Y}_{a}$ quadrature gives better results than using either $\hat{X}_{a}$ or $N_{a}$, which give oscillatory inferences. As it is, all of these give inferred uncertainties well above the SQL of $A=5.68\times10^{-18}$\;m for the parameters used. We note here that the actual calculated value of $\sigma(x)$, which oscillates between $1.025\times10^{-14}$\;m and $1.03\times10^{-14}$\;m,  is greater than the value inferred through measurement of the $\hat{Y}_{a}$ quadrature, which has a steady-state value of approximately $5\times10^{-16}$\;m. This is also the case with the inferred measurements of Ref.~\cite{MDR}, and is due to the almost perfect correlation between the position of the mirror and the $\hat{Y}_{a}$ quadrature.

\section{Nonclassical state preparation}
\label{sec:pensamentos}

The pendular cavity and variations on the theme have been proposed as useful devices for the preparation of nonclassical states of the cavity field and also of the mirror or mirrors. In this section we will give a brief review of some of these proposals, in chronological order, and explain why our results lead us to believe that they may not be as practical as the original authors suggest.

Bose {\em et al.\/}~\cite{Bose} suggest that this system may be used to prepare multi-component Schr\"odinger cat states of the field, near number states, and entangled states when two or more modes interact with the mirror, as well as Schr\"odinger cat-like states of the mirror prepared via quadrature measurements of the field. Beginning with a simple Hamiltonian without any pumping or dissipation terms, they develop a time evolution operator which is seen to have a Kerr-like term, which leads them to believe that this system may exhibit some of the nonclassical features of a Kerr interaction. They then assume that both the cavity field and the mirror are initially in coherent states. While this simplifies the mathematics, and is reasonably accurate for the cavity field, our calculations have shown that a coherent state is not a reasonable initial condition for the mirror.  As the nonclassical features predicted depend on this initial condition, which at one stage of the paper is taken to be a vacuum state (for simplicity), and interaction of the mirror with the environment is considered to be equivalent to that of a cavity field with a zero temperature reservoir, our results lead us to believe that these predicted nonclassical states will be, at the very least, extremely difficult to observe. 

In a subsequent article~\cite{macroknight}, the same authors propose that the pendular cavity may be used to probe the decoherence of a macroscopic object, namely the mirror. In this work the authors accept that the mirror will begin in a thermal state and note that a mixture of Schr\"odinger cat states of the mirror can therefore be produced by the interaction with the field. It is assumed that a superposition of the Fock states $|0\rangle$ and $|n\rangle$ can be prepared inside the cavity, with the most simply prepared value of $n=1$. Using the P-representation expansion of the density matrix in terms of coherent states, an expression is found for the time development of this density matrix. The field can then be measured via the interaction with a ground state two-level atom which is passed through the cavity. The probability of the atom being excited can be related to the decoherence rate of the spatially separated superposed coherent states of the mirror's motion. Examining this scheme with our parameters, we find that, with $n=1$, the spatial separation between two of the superposed coherent states has a maximum of $\Delta x_{max}=1.4\times10^{-22}$\;m, which is less than the thermal de Broglie wavelength, $\lambda_{dB}=2.19\times10^{-21}$\;m, so that it becomes difficult to claim we have a spatially separated superposition. When we take into account that the decomposition of the thermal state will include a huge number of coherent states, all with different phases, it seems that the practical realisation of this scheme may be more demanding technologically than the authors had supposed.

Another proposal uses radiation pressure to entangle two macroscopic mirrors~\cite{camerino}, but relies on Langevin equations which are linearised around their classical steady-state solutions. As we have shown above, this process is of rather doubtful validity when the mirrors are coupled to thermal reservoirs. A further idea is to produce an Einstein-Podolsky-Rosen (EPR) state in the position and momentum of two spatially separated oscillating mirrors using the output of a nondegenerate optical parametric oscillator (OPO)~\cite{zhang}. We note here that, although the authors call the oscillator an optical parametric amplifier, the nonlinear crystal is inside a pumped cavity, so we will follow the usual terminology. (See their Fig.2) Unlike most other treatments, this work uses an effective linear coupling between the light and the mirror. Like most of the others, the authors linearise Langevin equations around their steady state solutions. The OPO is treated via two-mode equations which do not describe the normal well-known threshold behaviour of such a device at all, and lead to the prediction of an entangled state of two combined quadratures, which is said to demonstrate an EPR correlation.  The two output fields of the OPO are used to drive the two mirrors, which drives them into an EPR state of position and momentum. The mechanical damping of the mirrors introduces a noise term $V_{b}=1+2n_{T}$, where $n_{T}=\coth(\hbar\omega_{m}/2k_{B}T)$ is the mean thermal phonon number. This noise term seems to be the quadrature variance assuming a number state $|n_{T}\rangle$, of the mirror. For the parameters we use, this expression gives a value of $V_{b}=6.7\times10^{6}$, several orders of magnitude lower than our estimate in Eq.~\ref{eq:VXYbthermal}, which gives $V_{b}=2.2\times10^{10}$. Due to the issues we have raised here, we are lead to believe that the EPR state of the two mirrors may not be as easily demonstrated as indicated by the authors. Mancini {\em et al.\/}~\cite{entrelacado} have also put forward a proposal to entangle two movable mirrors which form part of a four-mirror ring cavity. This work has the advantage of a description for the Brownian motion that is consistent with quantum mechanics at all temperatures, but the results presented are also obtained following a linearisation process, which we have shown to not be valid for reasonable temperatures. The final proposal we will consider here is by Marshall {\em et al.\/}~\cite{nakedemperor}, which treats the creation of superposition states of a macroscopic mirror ($\approx10^{14}$ atoms) via the interaction with a single photon. This work again uses a Hamiltonian approach without dissipation and assumes that the mirror can be prepared in its ground state, $|0\rangle$, which then allows for analytical solutions. The inclusion of decoherence and finite temperature follow the approach of Ref.~\cite{macroknight} and we therefore have the same doubts about the physical viability of this proposal.
           
\section{Cooling by feedback}
\label{sec:realimentacao}

It may be thought that to reduce the thermal noise of the mirror it is sufficient to cool to a lower temperature and impressive reductions in thermal noise, of the order of $10^{3}$, have been achieved~\cite{Pinard,exfeed}. However, it seems that the method of feedback cooling leads to a thermal equilibrium of the mirror at a lower temperature. For our system, using Eq.~\ref{eq:VXYbthermal}, a reduction by this factor leads to $V(\hat{X}_{b})=V(\hat{Y}_{b})\approx10^{7}$, which is still far above the coherent state level where linearisation of the equations may be expected to work. Ref.~\cite{nakedemperor} suggests that a mirror may be cooled to as low as $2$\;mK by dilution refrigeration, which would give us $V(\hat{X}_{b})=V(\hat{Y}_{b})\approx2.27\times10^{5}$, which, if we assume that feedback cooling from this temperature is as efficient as at room temperature, would allow the variances to be further reduced to something of the order of $10^{2}$. This is now actually smaller than the mean number of quanta, $|\beta|^{2}$, equal to $1.6\times10^{3}$, although the variance in the number will now be of the order of $10^{6}$, so that linearisation will still not be reliable. 

Standard cooling by feedback depends on the interaction of the electromagnetic field with the mirror phonons, which is proportional to $\hat{a}^{\dag}\hat{a}\hat{X}_{b}$, which will not drive the mirror toward a coherent state. A coupling term proportional to $\hat{a}^{\dag}\hat{b}-\hat{a}\hat{b}^{\dag}$ may be expected to do this, while squeezing of the mirror position would conceivably be possible with a coupling of the type $\hat{a}^{\dag\;2}\hat{b}-\hat{a}^{2}\hat{b}^{\dag}$ or $\hat{b}^{\dag\;2}\hat{a}-\hat{b}^{2}\hat{a}^{\dag}$. It is interesting to note that an effective coupling of the type required to drive the mirror toward a coherent state was used by Zhang {\em et al.\/}~\cite{zhang}, although the physical parameters used to develop the coupling Hamiltonian are far from those we have used in our analysis. As an example in that work, $\omega_{m}\gg\gamma$, whereas we have used the values $\omega_{m}=1.63\times10^{5}$\;s$^{-1}$ and $\gamma=3.14\times10^{6}$\;s$^{-1}$. Zhang {\em et al.\/} also consider that the mirror damping is of the amplitude form, which is not what we find using the Di\'osi master equation.
Mancini {\em et al.\/}~\cite{tartufo} have actually presented a feedback scheme based on an effective Hamiltonian which couples a light quadrature with $X_{b}$, later proposing that this method, which they call stochastic cooling, could be used to beat the SQL by achieving steady-state position squeezing of the mirror~\cite{Ribichini}. The effective coupling used again depends on the linearisation of the equations of motion. We feel that whether feedback can be used to cool a macroscopic oscillator towards, or even beyond, the SQL remains an open question and subject to further research. 

\section{Conclusion}

In conclusion, we have presented a fully quantum analysis of the pumped pendular cavity which does not depend on linearisation of the equations of motion. We have shown that the thermal noise of the mirror is overwhelming for typical temperatures and experimental parameters. This means that the linearisation procedure commonly used is of doubtful validity. Unless ways can be found to change the quantum state of the mirror in respect to the thermal excitations, we expect that the nonclassical states predicted in a number of theoretical analyses will not be able to be demonstrated experimentally. Feedback cooling techniques, while able to lower the effective temperature to an impressive degree, serve merely to produce another thermal state of the mirror at a lower temperature. The noise is still overwhelming if we wish to reach the SQL, or even beat it, which would be necessary for the detection of gravity waves. This is also the case if we wish to observe some of the quantum superpositions and entanglement which have been theoretically predicted.  

\section*{Acknowledgments}

This research was supported by the New Zealand Foundation for
Research, Science and Technology (Grant No. UFRJ0001), the Australian
Research Council, and the Brazilian agencies CNPq (Conselho Nacional de Desenvolvimento Cient\'{\i}fico e
Tecnol\'ogico) and CAPES (Coordena\c{c}\~ao de Aperfei\c{c}oamento de Pessoal de N\'{\i}vel Superior). Murray Olsen would like to thank Stojan Rebi\'c for interesting discussions and for his careful reading of the text. 


\end{document}